\documentclass[eqsecnum,aps,showpacs]{revtex4}
\begin{document}

\title{\bf Rejection Properties of Stochastic-Resonance-Based\protect\\
Detectors of Weak Harmonic Signals}
\author{R. P. Croce,  Th. Demma, V. Galdi, V. Pierro, I. M. Pinto}
\affiliation{University of Sannio at Benevento, IT} 
\author{F. Postiglione}
\affiliation{$D.I.^3E.$ University of Salerno, IT}

\def\beq{\begin{equation}}   
\def\eeq{\end{equation}}


\begin{abstract}
In (V. Galdi et al., Phys. Rev. {\bf E57}, 6470, 1998)
a thorough characterization 													
in terms of receiver operating characteristics (ROCs)
of stochastic-resonance (SR)  detectors 
of weak harmonic signals of known frequency 
in additive gaussian noise was given.									
It was shown that strobed sign-counting based
strategies can be used to achieve a nice trade-off
between performance and cost, by comparison
with non-coherent correlators.
Here we discuss the more realistic case where
besides the sought signal (whose frequency is
assumed known) further unwanted spectrally nearby
signals with comparable amplitude are present.
Rejection properties are discussed in terms of
suitably defined false-alarm and false-dismissal
probabilities for various values of interfering
signal(s) strength and spectral separation.
\end{abstract}

\pacs{05.40.-a, 02.50.-r, 84.40.Ua}
\maketitle


\section{- Background.}
~~~~Stochastic resonance (henceforth SR)
is a peculiar phenomenon observed in
a wide variety of physical systems \cite{footnote1}
acted by a mixture of a time-harmonic signal and 
(white, gaussian) noise, whereby 
the output spectral 
amplitude at the signal frequency shows a
{\em non-monotonic} dependence  {\em i)} 
on the noise strength at fixed signal
frequency and amplitude, and
{\em ii)} on the signal frequency, 
at fixed signal and noise levels.

The possible use of SR in connection with
{\em weak} signal detection experiments has
been repeatedly suggested.

The only meaningful comparison 
between different detection strategies is to compare
the related detection probabilities at the same
level of false alarm probabilities and available
signal-to-noise ratios. In this connection,
as demonstrated  in \cite{GalPP} and \cite{IncBul},
SR based detectors  do  {\em not}  outperform
matched filters. 
In \cite{GalPP}  it was
further shown that using SR as a pre-processor (signal-to-noise
ratio enhancer)  does {\em not} improve the performance
of  a matched-filter detector.  

Nonetheless, strobed sign counting detectors
could be interesting 
as computationally cheap alternatives to
noncoherent correlators, as discussed in \cite{GalPP}
and summarized in the next section. 

\section{- A Strobed Sign Counting SR Detector.}

~~~~The  possibly simplest 
SR-paradigm is the  Langevin system \cite{SR} :
\beq
\left\{
\begin{array}{l}
\dot{x}=-\displaystyle{\frac{dV(x)}{dx}}+A\sin(\omega_s t+\phi)+\epsilon n(t),\\
\\
x(0)=x_0,
\end{array}
\right.
\label{eq:Lang}
\eeq
with quartic symmetric potential:
\beq
V(x)=-a\frac{x^2}{2}+b\frac{x^4}{4},~~~a,b>0,
\eeq
where $n(t)$ is a stationary, zero mean, white Gaussian noise, with
autocorrelation\linebreak
\mbox{$E[n(t)n(t+\tau)]=\delta(\tau)$}.\\
The  probability density function
(henceforth PDF)  of   $x(t)$  in  (\ref{eq:Lang}),
denoted as $p(x,t)$ is ruled by the Fokker-Planck equation:
\beq
\left\{
\begin{array}{l}
\displaystyle{
\frac{\partial p(x,t)}{\partial t}}=
\displaystyle{\frac{\partial}{\partial x}}
\left\{
\left[
\displaystyle{\frac{dV(x)}{dx}}-A\sin(\omega_s t+\phi)
\right] p(x,t)
\right\}+
\displaystyle{\frac{\epsilon^2}{2}\frac{\partial^2 p(x,t)}{\partial x^2}},\\
~\\
p(x,0)=\delta(x-x_0).
\end{array}
\right.
\label{eq:FP}
\eeq
The solution of (\ref{eq:FP}), in the
absence of a signal $(A=0)$ is an even
function of $x$ \cite{GalPP}. 
In the presence of a signal, even-symmetry
is broken, and the asymmetry is maximum at 
\beq
t=t_k=\omega_s^{-1}
\left(
\frac{2k+1}{2}\pi-\psi
\right), ~~~~~k=0,1,2,..., 2N-1
\label{eq:strob_tim}
\eeq
where  $\psi$ is a (known \cite{GalPP}) 
phase-lag  introduced by the SR processor.

Symmetry breaking of the output  PDF
is perhaps \cite{footnote2}
the most natural signature of $A \neq 0$
in (\ref{eq:Lang}).
In \cite{GalPP} we  gave a thorough evaluation
in terms of receiver (detection) operating characteristics  
(henceforth ROCs) 
of the possibly simplest non-parametric detector 
based on  the above symmetry-breaking,
where from the output samples 
(\ref{eq:strob_tim}),  
one forms the time series
\beq
x_k=(-)^k x(t_k),
\eeq
and compares
\beq
N_+=\sum_{k=0}^{2N-1} U(x_k)
\label{eq:known_phi}
\eeq
where $U(\cdot)$ is Heaviside's step function,
to a suitable threshold $\Gamma$.
The above  will be henceforth referred
to as strobed sign-counting  
stochastic resonance detector (SSC-SRD).

The SSC-SRD performance is described
by the following false-alarm and false
dismissal probabilities \cite{GalPP}:
\beq
\alpha=\mbox{Prob}
\left\{
N_+ > \Gamma | \mbox{no signal}
\right\}=
I_{1/2}(\Gamma+1,2N-\Gamma),
\eeq
$${} $$
\beq
\beta=\mbox{Prob}
\left\{
N_+\leq \Gamma | \mbox{signal}
\right\}=
1-I_{P_+}(\Gamma+1,2N-\Gamma),
\label{eq:ROCs}
\eeq
where $P_+=\mbox{Prob}(x_k >0)$ and $I_p(x,y)$
is the incomplete Beta function .
The related ROCs
are typically worse by $\approx 3dB$ as compared 
to those of the matched filter \cite{GalPP}.

Generalization to the case 
where the initial phase of the sought
signal is unknown is straightforward, by letting:

\beq
N_+=\max_{m \in (0,N_s)}
\sum_{k=0}^{2N-1}
U\left[
(-)^k x
\left(
t_k+\frac{mT_s}{2N_s}
\right)
\right],
\label{eq:pick_max}
\eeq
where $T_s=2\pi/\omega_s$.
The resulting  unknown-initial-phase  detector
for  $N_s \stackrel{>}{\textstyle{\sim}} 10$
has nearly {\em the same} performance as  (\ref{eq:known_phi}),
which applies to the coherent (known initial phase)
case \cite{GalPP}, and is accordingly comparable
to that of the {\em noncoherent} correlator 
(std. optimum   benchmark detector for signals 
with unknown initial phase).

On the other hand, the SSC-SRD is computationally 
{\em extremely cheap}, requiring only binary and/or integer 
arithmetics, and thus quite appealing.

The obvious question is now related to the
{\em rejection properties} of the above detector, namely
to its ability to {\em discriminate} between spectrally
nearby signals (sought and unwanted) with comparable
amplitudes.

Note in this connection  that the frequency response
of the SSC-SRD  depends very little on the 
{\em stochastic resonance} condition, 
as shown in Fig. 1 (see \cite{GalPP}), 
where the steady-state value of $\max_t[P_+(t)]=\max_t[\mbox{Prob}(x(t) >0)]$
is displayed for several values of the $SNR$
as a function of  $\bar{\omega}_s=\omega_s T_k$,
where $\omega_s$ is the signal angular frequency and 
$T_k=\sqrt{2}\pi\exp(2\bar{V}_0)/a$ is the Kramers time.
In Fig. 1 $\bar{V}_0=a^2/(4b\epsilon^2)$ is 
the normalized potential-barrier height. 

A  numerical investigation of this issue will be the subject of the next Section.
\section{- ROCs in the Presence of Nearby Signals.}
~~~~In order to evaluate the performance of  the above
described SSC-SRD in the presence of  spectrally nearby
signals with comparable amplitude, we introduce the
following dimensionless parameters:
\beq
\rho_S=\frac{A_u}{A_w},~~~
\gamma=
\left|
\frac{\omega_{w}-\omega_{u}}{\omega_{w}}
\right|,~~~
\Delta\phi=\phi_{w}-\phi_{u},
\eeq
representing the unwanted-to-wanted (sought) signal-to-signal ratio ($SSR$),
the (scaled) frequency difference, and the phase lag,
where the suffixes $w$ and $u$ refer to the wanted and
unwanted signal, respectively.
More or less obviously, we also define the false-alarm ($\alpha)$
and false-dismissal  ($\beta$) probabilities as follows:
\beq
\alpha=\mbox{Prob}\left\{
N_+ > \Gamma |
s_{w}(t)=0~and~s_{u}(t) \neq 0
\right\},
\eeq
\beq
\beta=\mbox{Prob}\left\{
N_+ < \Gamma |
s_{w}(t) \neq 0 ~and~s_{u}(t) \neq 0
\right\}.
\eeq
Representative numerical  simulations \cite{Post}
are accordingly summarized in Fig.s  2 to 5 below.

Figure  2   describes a situation where $SNR=3$,
and the  unwanted signal frequency is pretty close 
to the sought one ($\gamma=10^{-4}$).
Only two detection characteristics are displayed,
corresponding to $\rho_S=2,3$
for the sake of readability, together with the limiting
curve corresponding to the absence of the unwanted signal (dashed line).
It is seen that the detector's
performance is only slightly deteriorated due to the presence
of the unwanted signal.

In Fig.  3 the $SSR$ is fixed at ($\rho_S=3$), 
and two different (normalised) frequency separations
$\gamma=10^{-2},~10^{-4}$ are considered. The limiting curve
corresponding to the absence of the unwanted signal (dashed line) is also included.
Again, the detector's performance is not appreciably spoiled.

In Fig. 4 the unwanted signal amplitude and spectral separation
are fixed, $\rho_S=3$ and $\gamma=10^{-4}$, and two values
of the phase lag are considered, $\Delta\phi=0,~\pi$.  
Once more, the detector's performance is negligibly
affected.

Finally,  Fig. 5  shows the key role played by 
the  number of strobed samples
used in (\ref{eq:pick_max})
at fixed  $\rho_S=3$, $\gamma=10^{-4}$, $\Delta\phi=0$ on the
SSC-SR detector's  rejection performance. 
It is seen that the unwanted signal is
rejected for $N > 10^3$, for which
the time series length corresponds to the spectral width required to
separate the wanted and unwanted signals. 
This shows that the rejection properties
are essentially related to the spectral filtering 
inherent to the strobing process.

\section{- Conclusions and Recommendations.}
~~~~Numerical simulations suggest that
strobed-sign-counting  stochastic-resonance-based detectors 
besides being computationally cheap, and nearly
as much performing as the standard non-coherent correlator,
do display nice rejection properties in the presence of 
spectrally nearby signals with comparable amplitudes.

Further study is under way to evaluate the potential 
of SSC-SR detectors in connection, e.g., with the search 
of weak nearby quasi-monochromatic signals in the
context of the search of gravitational waves. 


\section*{Acknowledgements}
~~~~The Authors thank prof Stefano Vitale (University of Trento) 
and dr. Stefano Marano (University of Salerno) for having posed 
the problem.  This work has been sponsored in part by the 
European Community through a Senior Visiting Scientist Grant
to I.M. Pinto at NAO - Spacetime Astronomy Division, Tokyo, JPN,
in connection with the TAMA project. I.M. Pinto wishes to thank
all the TAMA staff at NAO, and in particular prof. Fujimoto
Masa-Katsu and professor Kawamura Seiji for kind hospitality
and stimulating discussions.



\newpage

\begin{center}
\Large{Captions to the Figures}
\end{center}
\normalsize
Fig. 1 - Frequency response of SSC-SR detector for several values of SNR. Dashed curves refer
to the adiabatic approximation (quoted from \cite{GalPP}).\\
$$~$$ 					
Fig. 2 - ROCs of SSC-SR detector. $SNR=3$, $\gamma=10^{-4}$, $\Delta\phi=0$, $N=100$. 
The $\rho_S=0$ curve is shown dashed.\\
$$~$$
Fig. 3 -  ROCs of SSC-SR detector. $SNR=3$, $\rho_S=3$, $\Delta\phi=0$, $\gamma=10^{-2},10^{-4}$, $N=100$. 
The $\rho_S=0$ curve is shown dashed.\\
$$~$$
Fig. 4 -  ROCs of SSC-SR detector. $SNR=3$, $\rho_S=3$, $\gamma=10^{-4}$, $\Delta\phi=0,\pi$, $N=100$. 
The $\rho_S=0$ curve is shown dashed.\\
$$~$$
Fig. 5 -  ROCs of SSC-SR detector.  $SNR=3$, $\rho_S=3$, $\gamma=10^{-4}$, $\Delta\phi=0$, $N=50,100,1000$. 
The $\rho_S=0$ curve is shown dashed. Close-up in the inset.\\

\begin{thebibliography}{99}

\bibitem{footnote1}{See ,e.g., \cite{SR} for a survey of systems, signals and noises for which SR  is predicted/observed.}

\bibitem{SR}{L. Gammaitoni et al., Rev. mod. Phys., {\bf 70}, 223 (1998).}

\bibitem{GalPP}{V. Galdi et al., Phys. Rev. {\bf E57}, 6470 (1998).}

\bibitem{IncBul}{M.E. Inchiosa and A.R. Bulsara, Phys. Rev. {\bf E53}, R20 (1996).}

\bibitem{footnote2}{
Another possible signature is the
change in the distribution of residence-times 
in the potential wells (A.R. Bulsara et al., Phys. Rev. {\bf E67}, 016120 (2003)).}

\bibitem{Post}{F. Postiglione, Thesis, University of Salerno,  Jan. 1999.}


\end{thebibliography}
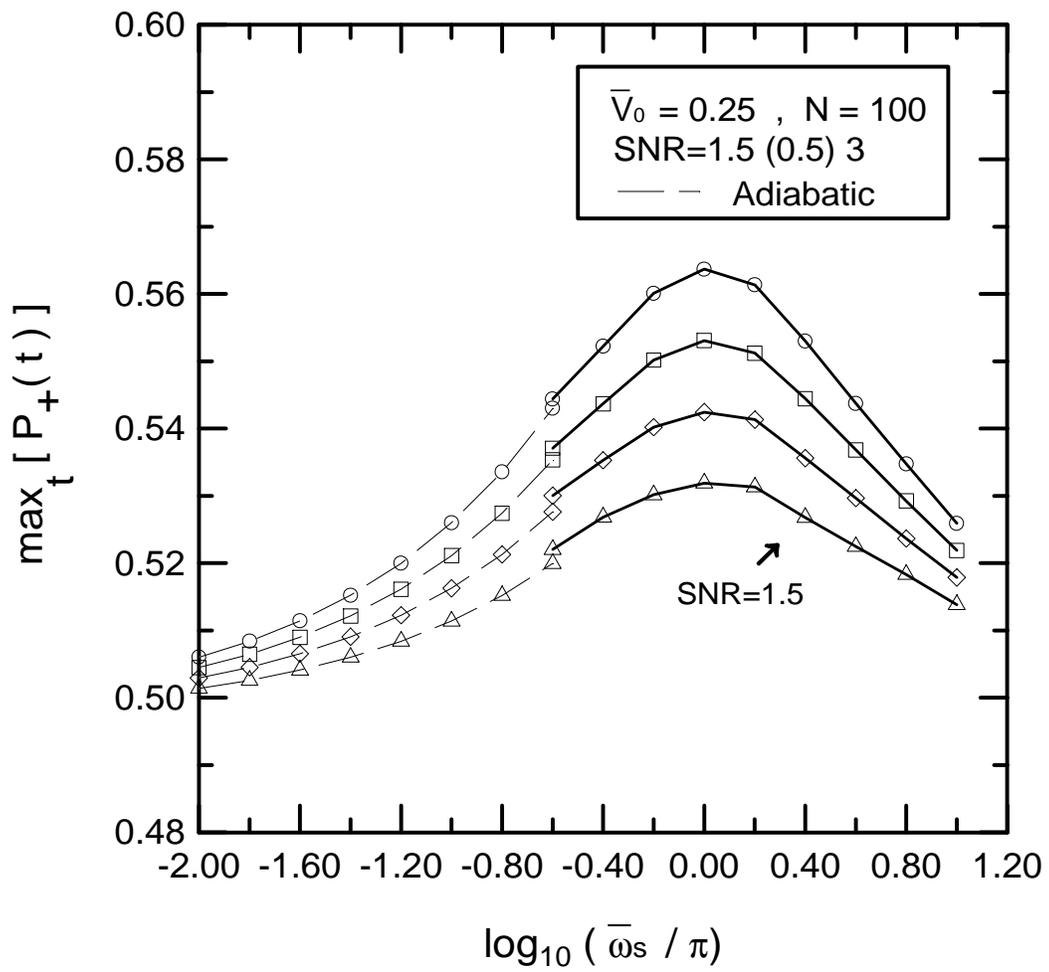
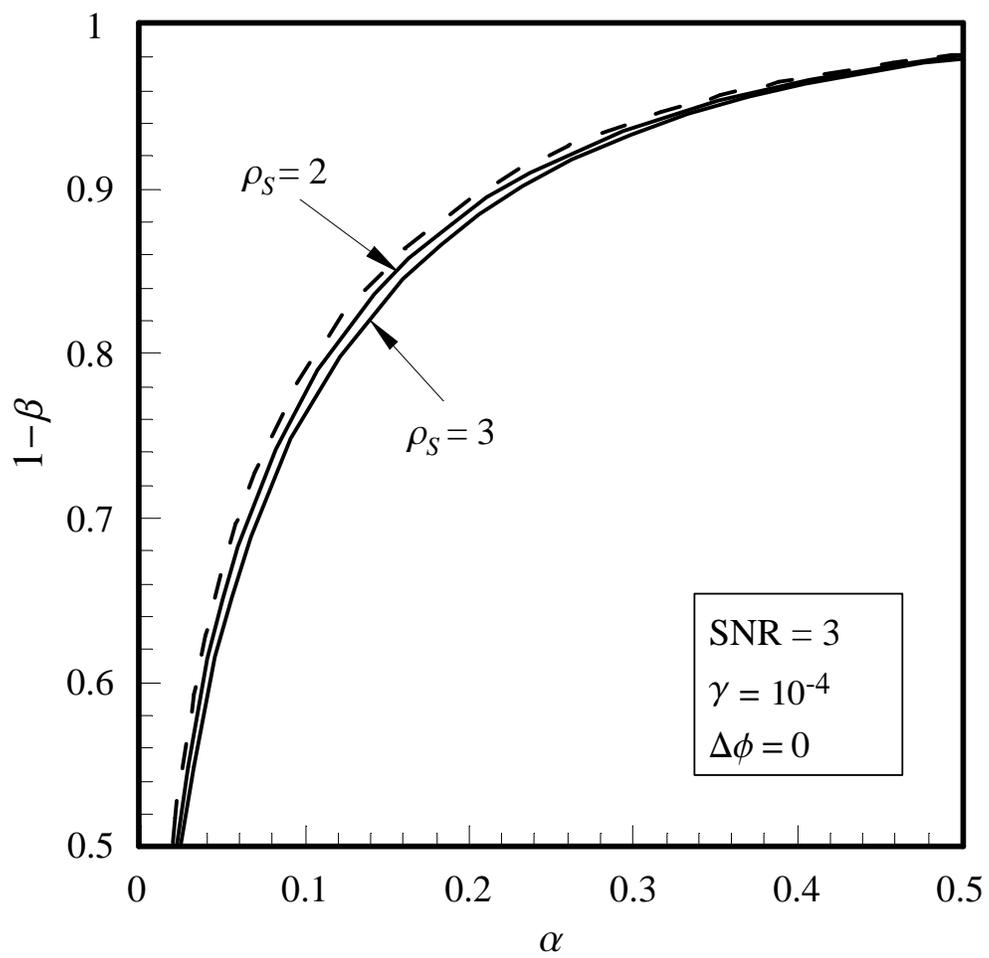
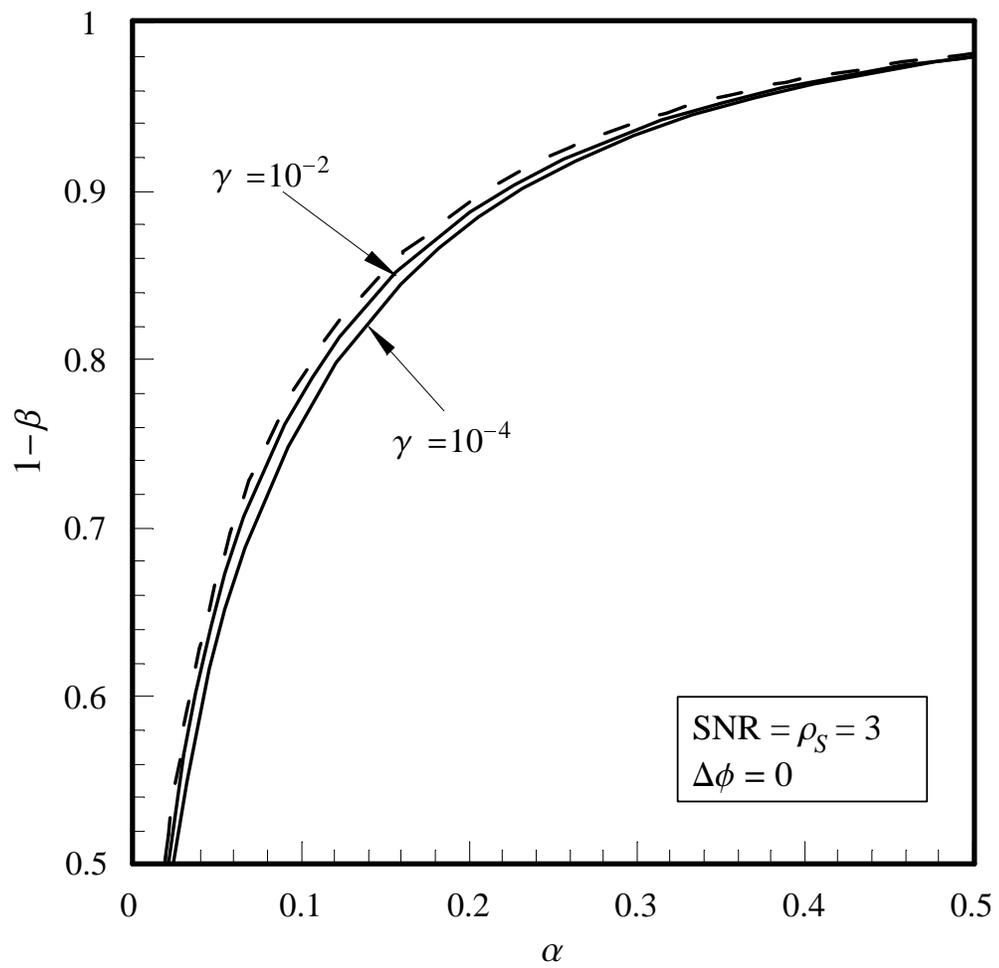
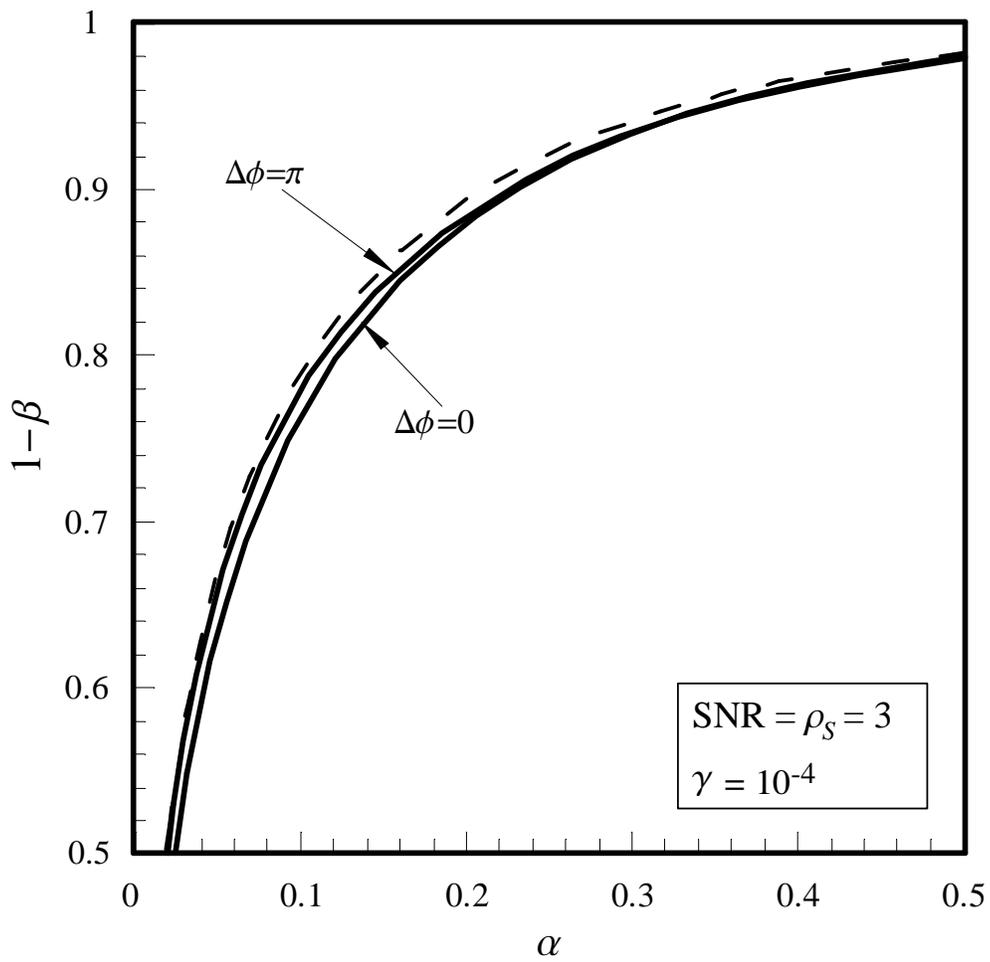
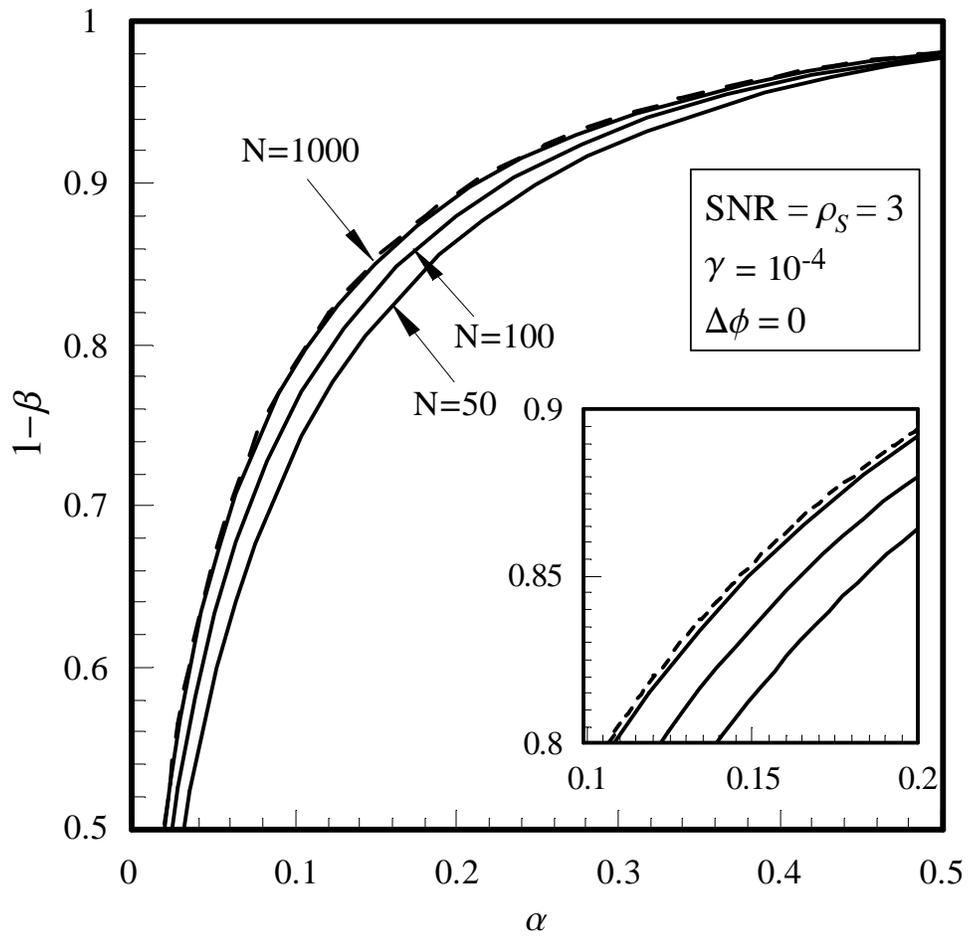
\end{document}